\documentclass[aps,preprint]{revtex4}
\usepackage{amsfonts}
\usepackage{amsmath}
\usepackage{amssymb}
\usepackage{color}
\usepackage{ulem}
\usepackage{graphicx,subfigure,epsfig}
\usepackage{epstopdf}
\usepackage{float}
\usepackage{booktabs}
\makeatletter

\makeatletter

\newcommand{\Rmnum}[1]{\uppercase\expandafter{\romannumeral #1}}

\begin{document}
	
\title{Quasinormal modes of Schwarzschild-like black hole surrounded by the pseudo-isothermal dark matter halo}

\author{Yanqiang Liu$^{1}$\footnote{e-mail:liuyanqiang@stu.scu.edu.cn} and Benrong Mu$^{2}$\footnote{e-mail: benrongmu@cdutcm.edu.cn}, Jun Tao$^{1}$\footnote{e-mail: taojun@scu.edu.cn},Yuling Weng$^{1}$\footnote{e-mail: wengyuling@stu.scu.edu.cn}}

\address{$^{1}$Center for Theoretical Physics,College of Physics, Sichuan University, Chengdu,610065,China\\
	$^{2}$Center for Joint Quantum Studies, College of Medical Technology, Chengdu University of Traditional Chinese Medicine, Chengdu, 611137, PR China}

\begin{abstract} 
The merger of binary black holes produces a series of decaying oscillations, during which energy is radiated in gravitational waves. The characteristic signal in the ringdown phase can be described by complex oscillation frequencies called quasinormal modes (QNMs). In this paper, we investigate the ringdown spectrum resulting from scalar field perturbations of black holes surrounded by pseudo-isothermal dark matter halos. 
The complex frequencies of these quasinormal modes are numerically computed using the sixth-order WKB approximation. Additionally, the time evolution of the scalar perturbations is examined using the finite difference method, considering various multipole numbers and dark matter halo parameters. For a static, spherically symmetric black hole, the photon sphere—composed of circular null geodesics—plays a crucial role in analyzing the black hole shadow. Furthermore, the connection between the black hole shadow and QNMs is explored in the eikonal limit.

\end{abstract}

\maketitle

\section{Introduction}

Einstein's theory of general relativity revolutionized our understanding of spacetime and gravitation, predicting the existence of black holes. Many galaxies host supermassive black holes at their centers, whose immense gravity prevents light from escaping, making direct observation challenging. The Event Horizon Telescope (EHT) first captured the shadow of the black hole at the center of the M87 galaxy, aligning with general relativity's predictions \cite{EventHorizonTelescope:2019dse}. 
These observations provide new ways to constrain black hole models, and the study of black hole shadows has garnered significant attention, particularly in different gravitational backgrounds~\cite{Ling:2021vgk, Yang:2022btw, Vagnozzi:2022moj, Jusufi:2019nrn,  Haroon:2018ryd, Bambi:2019tjh, Vagnozzi:2019apd, Kumar:2020yem, Ghosh:2020ece,Adler:2022qtb}. 

The center of galaxies often exhibits complex distributions of material fields, such as accretion disks, galactic nuclei, strong magnetic fields, and dark matter halos.
Interactions between a black hole and its surrounding environment can induce perturbations in the black hole, leading to the emission of gravitational waves. These gravitational waves are divided into three phases in their temporal evolution, with the third phase featuring quasinormal modes (QNMs) with complex frequencies~\cite{Konoplya:2011qq, Vishveshwara:1970zz}. The real part of these complex frequencies represents the oscillation frequency of the perturbed black hole, while the imaginary part indicates the decay rate. Population studies have investigated QNMs in different black hole spacetime\cite{Sun:2023woa, Zhang:2021bdr, Hendi:2020zyw, Jafarzade:2020ova, Anacleto:2021qoe, Lambiase:2023hng,Ghosh:2022gka, Pedrotti:2024znu,Ghosh:2023etd,
Jusufi:2019ltj, Hamil:2024ppj}. Also, various methods for computing black hole QNMs have been developed ~\cite{Schutz:1985km, Iyer:1986np, Ferrari:1984zz, Churilova:2021nnc, Leaver:1985ax, Dolan:2007mj}. 
The gravitational wave detectors like LIGO, VIRGO, and LISA can detect signals from black holes, and the primary contributors to these signals would be the fundamental modes of the QNMs. This is one of the reasons why people are particularly interested in QNMs~\cite{Aso:2013eba, Liu:2020eko, Ruan:2018tsw}.

The nature of dark matter has long intrigued scientists, leading to considerable speculation and making it a key topic in cosmology~\cite{deSwart:2017heh}. In 1933, F. Zwicky first provided evidence for dark matter by measuring the redshift of galaxies in the Coma cluster~\cite{Zwicky:1933gu}. Over the past few decades, cosmologists have developed a standard model that describes the large-scale structure of the universe, the Lambda Cold Dark Matter ($\Lambda$CDM) model. It not only explains the universe evolved, but also galaxies and galaxy clusters formed~\cite{Navarro:1995iw, Navarro:1996gj}. N-body simulations have shown that a cold dark matter (CDM) halo with a Navarro-Frenk-White (NFW) profile exhibits a universal, spherically averaged density profile~\cite{Jusufi:2019nrn}. However, the CDM model faces challenges at smaller scales, particularly at the galactic level, where it fails to fully describe the small-scale structure of the universe~\cite{Robles:2012uy}. Observations of the rotation curves of dwarf and low-surface-brightness (LSB) galaxies over the past few decades have shown that the inner regions have flatter cores rather than the density cusp predicted by the cold dark matter \cite{McGaugh:1998tq}. To address this discrepancy, numerous modifications and alternatives to the standard CDM paradigm have been proposed~\cite{Spergel:1999mh,Kamionkowski:1999vp}.
Studies have also shown that supermassive black holes at the centers of galaxies can significantly increase the dark matter density, leading to the formation of a ``spike''~\cite{Sadeghian:2013laa,Fields:2014pia,Zhao:2023tyo}. Therefore, studying black holes surrounded by dark matter can help us better understand the distribution of dark matter in the center of galaxies. A pseudo-isothermal profile halo, proposed by Begeman, assumes a spherically symmetric halo with a density profile that remains constant (isothermal) at the center and gradually decreases with distance from the center \cite{Begeman:1989kf}. This model avoids the central singularities seen in the NFW profile and is consistent with observations of LSB galaxies, where rotation curves either flatten or rise slightly at large radii.
We will focus on the pseudo-isothermal halos and derive a spherically symmetric static metric for a Schwarzschild-like black hole. As a commonly  used dark matter profile model, it provides an important theoretical framework for understanding dark matter on galaxy structure and dynamics through their feature of density.

In this paper, we employ the WKB approximation method to compute the QNMs of black holes surrounded by pseudo-isothermal dark matter halos. The paper is structured as follows: In Sect.~\ref{sect2}, we study the spacetime distribution of Schwarzschild-like black holes in pseudo-isothermal dark matter halos and derive the null geodesic equations and the shadow of a black hole. In Sect.~\ref{sect3}, we investigate the scalar field perturbations and the influence of various parameters on QNMs via sixth-order WKB approximate, while also evaluating their temporal evolution. In the eikonal limit, the connection of the shadow and the real part of QNMs has been considered. A comprehensive summary of our findings is presented in Sect.~\ref{sect4}.

\section{The shadow in pseudo-isothermal hole model }\label{sect2}
  In this section, we study the black holes immersed in pseudo-isothermal dark matter hole.
To derive the spherically symmetric black hole metric, we first consider the spacetime line elements associated with a pure dark matter halo. By solving the Einstein field equations, the resulting metric can describe a black hole surrounded by the pseudo-isothermal dark matter halo~\cite{Yang:2023tip}
\begin{align}\label{metriceq}
	d s^2=-f(r) d t^2+f(r)^{-1} d r^2+r^2\left(d \theta^2+\sin ^2 \theta d \phi^2\right),
\end{align}
where $f(r)$ has the form of the following,
\begin{equation}\label{meteq}
	f(r)=\left(r_0^2+r^2\right){}^{4 \pi  \rho _0 r_0^2} \text{exp}\left[\frac{8 \pi  \rho _0 r_0^3  \arctan
		\left(\frac{r}{r_0}\right)}{r}\right]-\frac{2M}{r}.
\end{equation}
Here $M$ is the mass of the black hole, $\rho_0$ denotes the central halo density and  $r_0$ denotes the halo core radius. These dark matter parameters roughly reflect the distribution of pseudo-isothermal halos throughout the entire galaxy. Additionally, if the existence of dark matter halos is not taken into account(i.e.$\rho_0=0$), the metric function $f(r)$ would revert to the classical Schwarzschild solution.
It's important to note that these parameters are typically obtained by fitting the observed data of rotation curves from different galaxies to the corresponding density profiles~\cite{Robles:2012uy}. In this article, we mainly use the data characteristic radius parameter $r_0=0.57 \,\mathrm{kpc}$ and the density parameter $\rho_0=0.0464 \, M_{\odot} / \mathrm{kpc}^3$ from the LSB galaxy ESO1200211\cite{Robles:2012uy, Lin:2019yux}. Moreover, the values of $r_0$ and $\rho_0$ may vary with different galaxies. Therefore, we can study the different values of dark matter parameters in this work. 
For convenience in subsequent calculations, the dark matter parameters in Eq.(\ref{metriceq}) are guaranteed to be in black hole units, which can be achieved through the following method
\begin{equation}
\begin{aligned}
	r_0=\frac{r_0}{r_{BH}},\quad \rho_0=\frac{\rho_0}{M_{BH}/(\frac{4}{3}\pi (r_{BH})^3)},
\end{aligned}
\end{equation}
where $r_{BH}=G M_{BH}/c^2$ represents the radius of black hole.

Due to the strong gravitational field near the black hole, photons are  forced to travel along circular null geodesics on the photon sphere, playing an important role in determining  size of the black hole shadow observed by distant observers.
We first study the motion of photons in the gravitational field near a black hole by utilizing the Lagrangian equations to derive the equations of motion. The corresponding Lagrangian is given by
\begin{align}
	2{\cal L}=\-f(r)\dot{t}^2 +\frac{1}{f(r)}\dot{r}^2+r^2\dot{\theta}^2+r^2\sin^2\theta \dot{\varphi}^2,
\end{align}
where the dot represents the derivative with respect to the affine parameter $\tau$. 
Then we can give an expression for the generalized canonical momenta
\begin{equation}\label{momentum eq}
	\begin{aligned}
		p_{t}&=-f(r)\dot{t}=-E,\\
	p_{\varphi}&=r^2\sin^2\theta\dot{\varphi}=L.
	\end{aligned}
\end{equation}
It's worth noting that the metric of the black hole surrounded by pseudo-isothermal dark matter halos spacetime has no connection with $t$ and $\varphi$. This implies that the spacetime structure possesses two Killing vectors. So we can obtain the conserved energy $E$ and momentum $L$ in Eq. (\ref{momentum eq}). 

Then we study the orbit of the photon sphere with the Hamilton-Jacobi equation, which is defined as
\begin{align}\label{hy}
	\frac{\partial\mathcal{S}}{\partial\tau}=-\mathcal{H},
\end{align}
where $\mathcal{S}$ denotes the Jacobi action, and $ \mathcal{H}$ denotes the Hamiltonian. By utilizing $ \mathcal{H}$ , we can rewrite Eq.(\ref{hy}) as
\begin{align}\label{HJe}
	\frac{\partial}{\partial\tau}\mathcal{S}=-\frac{1}{2}g_{\mu\nu}\frac{\partial\mathcal{S}}{\partial x^{\mu}}\frac{\partial \mathcal{S}}{\partial x^{\nu}}.
\end{align}
Using the method of separation of variables, the expression for the action $\mathcal{S}$ can be given by
\begin{align}\label{actionJ}
	\mathcal{S}=\frac{1}{2}m^2\tau-Et+\mathcal{S}_r(r)+\mathcal{S}_{\theta}(\theta)+L\varphi,
\end{align}
where $m$ represents the mass of the particle, $\mathcal{S}_r(r)$ and $\mathcal{S}_{\theta}(\theta)$ denote the radial and angular functions. Substituting Eq. (\ref{actionJ}) into Eq. (\ref{HJe}), we obtain 
\begin{equation}
		\dot{t}=\frac{E}{f(r)},\quad
		\dot{\varphi}=\frac{L}{r^2\sin\theta},\quad
		r^2\dot{r}=\pm\sqrt{\mathcal{R}},\quad
		r^2\dot{\theta}=\pm\sqrt{\mathcal{\varTheta}},
\end{equation}
where $\mathcal{R}$ and $\mathcal{\varTheta}$ are defined as
\begin{equation}
	\begin{aligned}
\mathcal{R}&=r^4E^2-r^2(\mathcal{K}+L^2)f(r),\\
\mathcal{\varTheta}&=\mathcal{K}-L^2\cot^2\theta.
	\end{aligned}
\end{equation}
Here $\mathcal{K}$ is defined as Carter’s separation constant. To study the shape of the black hole shadow, we consider the motion of photons on null geodesics. The two impact parameters indicate the motion of photons near the black hole
\begin{align}
	\xi=\frac{L}{E}, \quad \eta=\frac{\mathcal{K}}{E^2}.
\end{align}
\begin{figure*}
	\begin{tabular}{ccc}
		\begin{minipage}{0.5\hsize}
			\centering
			\includegraphics[width=5.5cm,height=5.5cm]{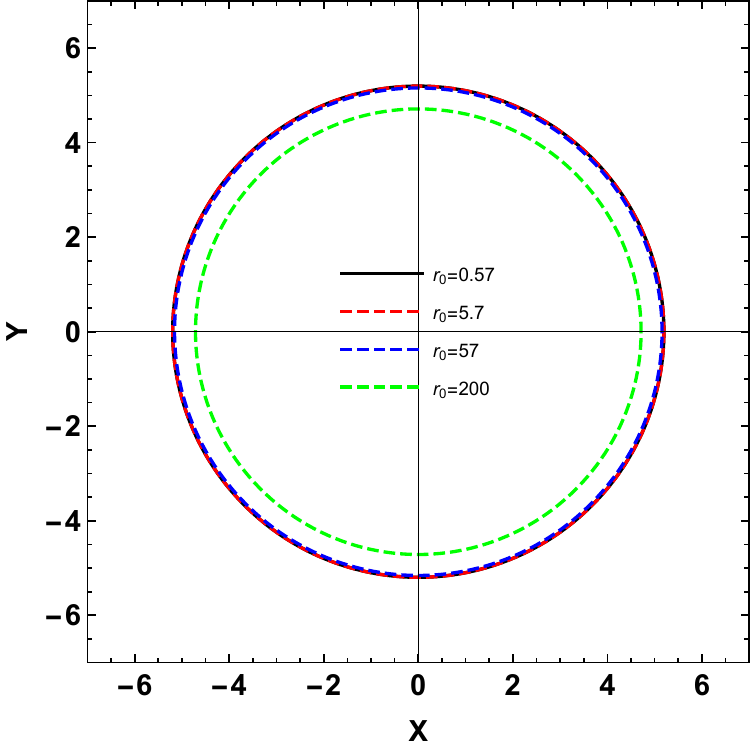}
		\end{minipage} &
		\begin{minipage}{0.5\hsize}
			\centering
			\includegraphics[width=5.5cm,height=5.5cm]{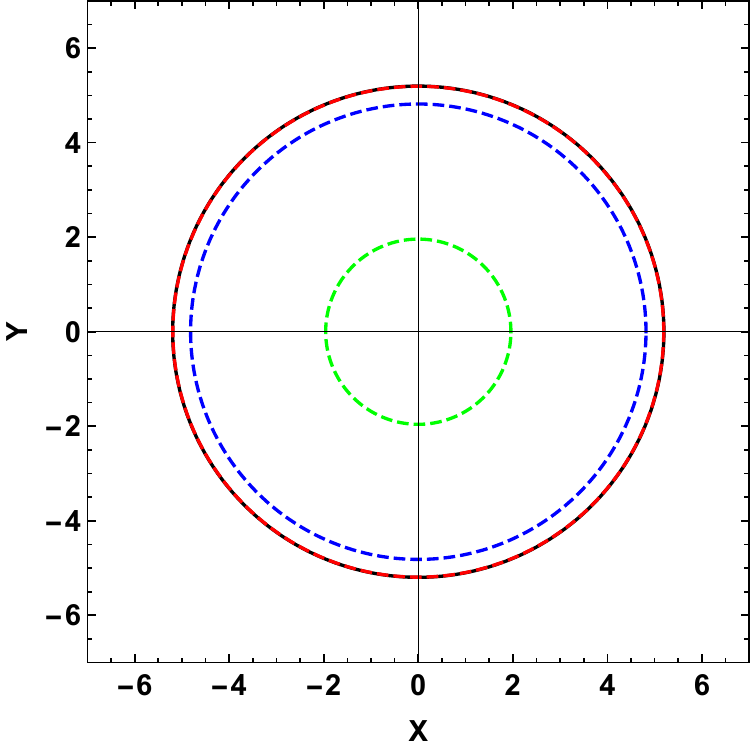}
		\end{minipage} 
	\end{tabular}
	\caption{The influence of dark matter parameters on the shadow shape of the black hole. The left panel is configured with parameter $\rho_0=0.0464\, M_{\odot} / \mathrm{kpc}^3$, while the right is configured with parameter  $\rho_0=0.464\, M_{\odot} / \mathrm{kpc}^3$.}
	\label{fig:ps1}
	
	\begin{tabular}{ccc}
		\begin{minipage}{0.5\hsize}
			\centering
			\includegraphics[width=5.5cm,height=5.5cm]{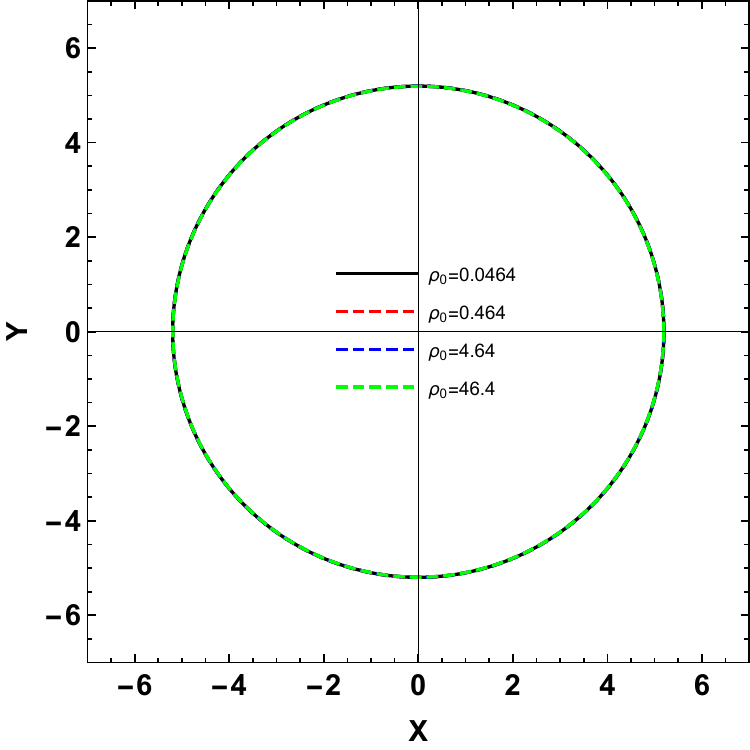}
		\end{minipage} &
		\begin{minipage}{0.5\hsize}
			\centering
			\includegraphics[width=5.5cm,height=5.5cm]{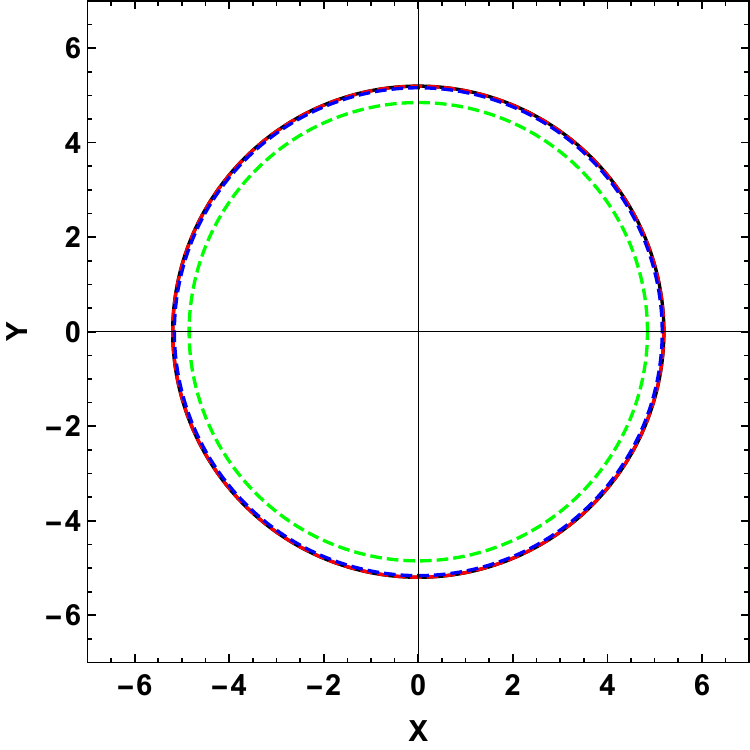}
		\end{minipage} 
	\end{tabular}
	\caption{The influence of dark matter parameters on the shadow shape of the black hole. The left panel is configured with parameter $r_0=0.57 \,\mathrm{kpc}$, while the right is configured with parameter  $r_0=5.7\, \mathrm{kpc}$. }
	\label{fig:ps2}
\end{figure*}
 The formation of the black hole shadow being related to unstable circular photon orbits. Imposing the following conditions determine the photon orbits
\begin{align}
	\mathcal{R}(r)|_{r_{ph}}=0,\quad \frac{d\mathcal{R}}{dr}\bigg|_{r_{ph}}=0,\quad \frac{d^2\mathcal{R}}{dr^2}\bigg|_{r_{ph}}< 0,
\end{align}
where $r_{ph}$ represents the radius of the photon sphere. The second derivative being less than zero indicates an unstable condition. According to the above conditions, the impact parameters can be rewritten as follows
\begin{align}
\xi^2+\eta=\frac{r_{ph}^2}{f(r_{ph})}.
\end{align}
 There is no analytical solution to the equation describing the photon sphere. Hence, we can employ numerical solving of the following equation to obtain the value of $r_{ph}$ and subsequently derive 
$\xi^2+\eta$ associated with it,
\begin{align}
	r_{ph}f'(r_{ph})-2f(r_{ph})=0.
\end{align}
The motion of photons inside a black hole, surrounded by dark matter halo, can be calculated based on the study of the preceding impact parameters. Now we introduce a celestial coordinate system to describe the actual black hole shadow observed by an observer~\cite{Johannsen:2013vgc, Hioki:2009na}
\begin{equation}
	\begin{aligned}
     X&=\lim_{r\to \infty}\Big(-r^2\sin\theta\frac{d\phi}{dr}\Big),\\
     Y&=\lim_{r\to \infty}\Big(r^2\frac{d\theta}{dr}\Big),
	\end{aligned}
\end{equation}
where $r$ represents the distance from the observer to the black hole, and $\theta$ represents the inclination angle of the black hole.  The simplified celestial coordinates are shown below for asymptotically flat black holes
\begin{equation}
	\begin{aligned}
		X&=-\frac{\xi}{\sin\theta},\\
		Y&=\pm\sqrt{\eta-\xi^2\cot^2\theta}.
	\end{aligned}
\end{equation}
Consider an observer located on the equatorial plane, that is $\theta=\pi/2$, the celestial coordinates can be expressed as
\begin{align}
	X^2+Y^2=\xi^2+\eta=R^2_{S},
\end{align}
where $R_{S}$ can present the shadow image of the black hole surrounded by the pseudo-isothermal halo.
 Utilizing astronomical coordinates can derive a black hole's shadow and investigate the effect of various parameters on its form. We investigated the influence of the hole core radius $r_0$ on the shadow, as shown in Fig.\ref*{fig:ps1}.  The left panel shows the black hole shadow when the central density $\rho_0=0.0464 \, M_{\odot} / \mathrm{kpc}^3$. The right panel shows the case where the central density $\rho_0=0.464\, M_{\odot} / \mathrm{kpc}^3$. Various colors of lines represent distinct values of the hole core radius parameter, which is essentially a standard circle.  It is shown that with increasing core radius $r_0$, the shadow gradually decreases. In the right panel, an obvious reduction in the shadow radius occurs when $r_0=200\, \mathrm{kpc} $. When the dark matter density is higher, the effect of the halo core radius on the black hole shadow becomes more significant.  We also investigated the result of the central density $\rho_0$ on the shadow, as shown in Fig.\ref*{fig:ps2}. Here we set the hole core radius  $r_0=0.57 \, \mathrm{kpc}$ for the left panel and  $r_0=5.7\, \mathrm{kpc}$ for the right panel, respectively. Various colors represent the influence of different central density parameters on the shadow of the black hole. With increasing central density $\rho_0$, the shadow of the Schwarzschild-like black hole also gradually diminishes.

\section{Quasinormal modes}\label{sect3}
\begin{figure*}
	\begin{tabular}{cc}
		\begin{minipage}{0.5\hsize}
			\centering
			\includegraphics[width=8.2cm,height=5.5cm]{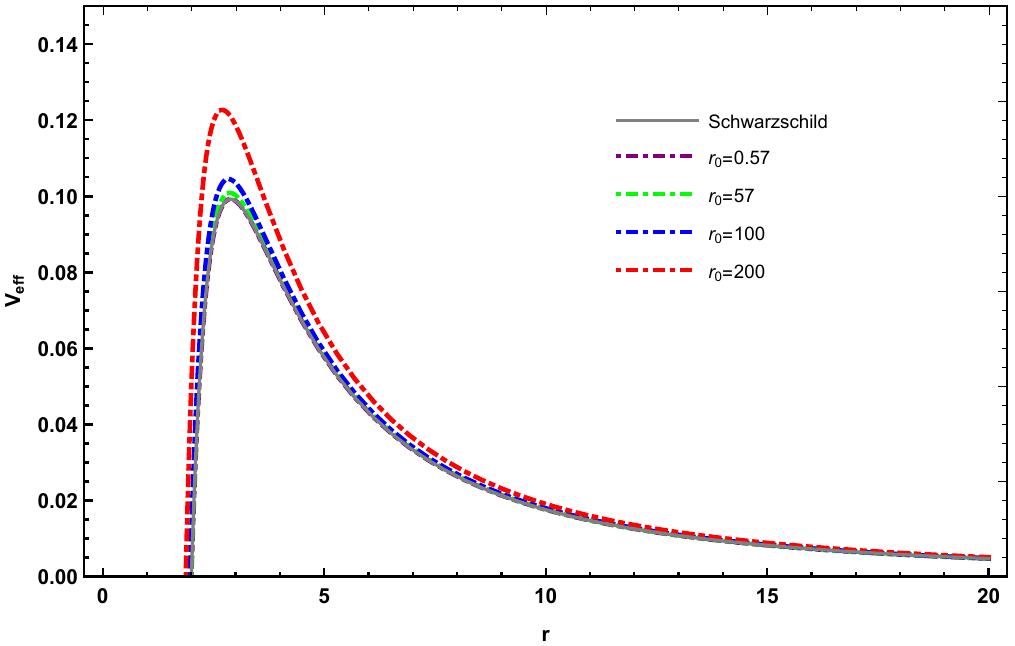}
		\end{minipage} &
		\begin{minipage}{0.5\hsize}
			\centering
			\includegraphics[width=8.2cm,height=5.5cm]{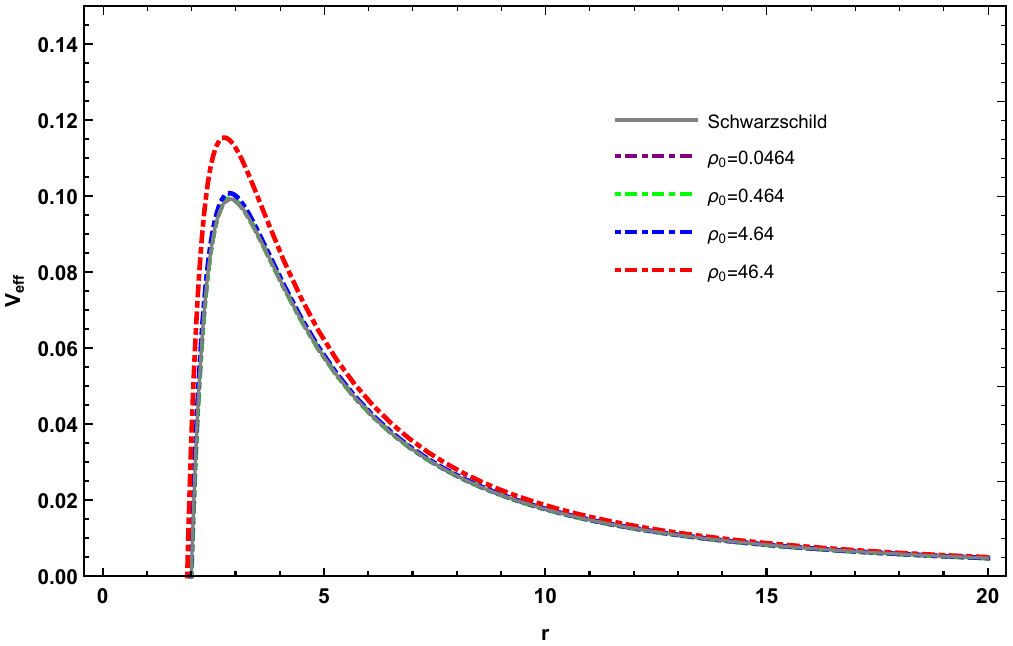}
		\end{minipage} 
	\end{tabular}
	\caption{Here we study the effect of different parameters $r_0$ and $\rho_0$ for the effective potential($l=1$). The left panel is $\rho_0=0.0464\, M_{\odot} / \mathrm{kpc}^3$ and the right panel is $r_0=5.7\, \mathrm{kpc}$.}
	\label{figVeff}
\end{figure*}

The quasinormal modes refer to oscillations with complex frequencies representing energy dissipation, which indicate some distinctive properties of black holes, such as mass, charge, angular momentum, and so forth. In this section, we discuss the scalar field perturbation of Schwarzschild-like BHs and search for the corresponding effective potential. The Klein-Gordon equation can be used to describe the motion of a particle in the scalar field,
\begin{align}\label{KGe}
	\frac{1}{\sqrt{-g}}\partial_{\nu}(g^{\mu\nu}\sqrt{-g}\partial_{\mu}\delta\phi)=0.
\end{align}
The scalar perturbation in the spherical symmetry of spacetime can be written as
\begin{equation}
	\delta\phi(t,r,\theta,\phi)=e^{-i\omega t}
	\frac{\Psi_{\omega l}(r)Y_{lm}(\theta,\phi)}{r},
\end{equation}
where $Y_{lm}(\theta,\phi)$ represents the spherical harmonics, $l$ takes natural numbers, $m$ takes integers, and $l>|m|$.
By substituting the above equation into Eq. (\ref{KGe}) and separating the angular variables, we can obtain a Schr\"{o}dinger-like equation
\begin{equation}\label{waveeq}
	\big(\frac{d^2}{d r_*^2}+\omega^2-V_{eff}(r)\big)\Psi_{\omega l}(r)=0,
\end{equation}
where the tortoise coordinate $r_*$ is defined as $d r_*/dr=1/f(r)$. And the effective potential is given by
\begin{align}
	V_{eff}=f(r)\Big(\frac{l(l+1)}{r^2}+\frac{f'(r)}{r}\Big).
\end{align}
It is worth noting that the form of the effective potential has a potential barrier in the tortoise coordinates. The effect of variations in two different dark matter parameters ($r_0$ and $\rho_0$) on the effective potential is illustrated in Fig.\ref{figVeff}. Both hole core radius and hole density have minimal effect on the effective potential at small values. The significant result on the effective potential will occur when the value of the parameter is large enough. We speculate that when either the fixed $r_0$ or $\rho_0$ value is small, the increase of the other parameter at small scales has a weaker manifestation on the potential barrier of the effective potential, meaning that the influence of dark matter halos on Schwarzschild black holes is not significant. Additionally, the relationship between quasinormal frequencies and the maximum of $V_{eff}$ suggests that variations in $r_0$ and $\rho_0$ may also have similar effects on quasinormal frequencies.
\subsection{WKB method}
Based on the effective potential obtained above, we can study the quasinormal frequencies of Schwarzschild-like black holes in pseudo-isothermal halos. One can calculate the quasinormal modes of this black hole using the WKB method and investigate its associated properties. It is important to note  that QNM emerges when appropriate boundary conditions are applied before performing the calculations,
\begin{equation}
	\begin{aligned}
		&\Psi_{\omega l}(r_*) \sim e^{-i\omega r_*}, \quad r_{*}=-\infty,\\
    	&\Psi_{\omega l}(r_*) \sim e^{i\omega r_*}, \quad  r_{*}=+\infty,
	\end{aligned}
\end{equation}
where purely ingoing modes occur at $r_{*} = -\infty $ (at the event horizon) and purely outgoing modes occur at $r_{*} = + \infty $ (at the spatial infinity). The spectrum expression in Eq. (\ref{waveeq}) cannot be obtained analytically. Therefore, we introduce a numerical method for calculating QNMs. This method, first proposed by Schutz and Will, was used to address black hole scattering problems~\cite{Schutz:1985km}. Later, further developments were made by Iyer, Will, and Konoplya~\cite{Iyer:1986np, Konoplya:2011qq}. The WKB method has evolved to the 13th order.  In this paper, we adopt the sixth-order WKB approximation while ensuring precision
\begin{align}
	\frac{i(\omega^2-V_0)}{\sqrt{-2V_0^{''}}}-\sum_{i=2}^{6}\Lambda_i=n+\frac{1}{2},\quad (n=0,1,2,\cdots)
\end{align}
where the superscript of $V_0^{''}$ denotes the second derivative over the tortoise coordinate $r_*$, $V_0$ represents the maximum value of the effective potential, and $\Lambda_i$ is the $i$th order revision terms depending on the values of the effective potential~\cite{Konoplya:2011qq}. 

	\begin{table}[h!]
		\centering
		\setlength{\tabcolsep}{4pt} 
		\renewcommand{\arraystretch}{0.4} 
		\begin{tabular}{|c|c|c|c|c|c|}
			\hline
			& Re $(\omega)$ & Im $(\omega)$  & &Re $(\omega)$ &Im $(\omega)$ \\
			\hline
			$l=1$ & 0.3294344 & $-0.0962563$ & $n=1$ & 2.0446636 & $-0.2861228$ \\
			\hline
			$l=2$ & 0.5063169 & $-0.0961233$ & $n=2$ & 2.0781986 & $-0.4691763$ \\
			\hline
			$l=3$ & 0.6917286 & $-0.0961482$ & $n=3$ & 2.1240966 & $-0.6426535$ \\
			\hline
			$l=4$ & 0.8801978 & $-0.0961714$ & $n=4$ & 2.1788972 & $-0.8054876$ \\
			\hline
			$l=5$ & 1.0700936 & $-0.0961866$ & $n=5$ & 2.2396675 & $-0.9577722$ \\
			\hline
			$l=6$ & 1.2607675 & $-0.0961964$ & $n=6$ & 2.3041843 & $-1.1002192$ \\
			\hline
			$l=7$ & 1.4519116 & $-0.0962030$ & $n=7$ & 2.3708640 & $-1.2337799$ \\
			\hline
			$l=8$ & 1.6433612 & $-0.0962076$ & $n=8$ & 2.4386159 & $-1.3594354$ \\
			\hline
			$l=9$ & 1.8350206 & $-0.0962110$ & $n=9$ & 2.5067034 & $-1.4780997$ \\
			\hline
			$l=10$ & 2.0268300 & $-0.0962134$ & $n=10$ & 2.5746345 & $-1.5905847$ \\
			\hline
		\end{tabular}
		\caption{The quasinormal modes frequencies of the black hole surrounded by the pseudo-isothermal dark matter halo in the scalar field. The left is the case where $n$ is fixed($n=0$), while the right is the case where $l$ is fixed($l=10$). The corresponding calculation parameter is $M=1$, $r_0=0.57 \,\mathrm{kpc}$ and $\rho_0=0.0464 \, M_{\odot} / \mathrm{kpc}^3$.}
		\label{fig:tablea}
	\end{table}

 At present, we delve into the quasinormal mode frequencies of black holes encompassed by pseudo-isothermal dark matter halos.
As dissipative systems, the oscillatory behavior of black holes can be calculated through pure ingoing waves at the horizon and pure outgoing waves at infinity. Without losing generality, we calculate the QNM frequency of a Schwarzschild black hole with different angular quantum numbers $l$ and overtone number $n$ in a massless scalar field. We divide these QNM frequencies into real and imaginary parts, where the real part represents the oscillation frequency of the system, and the imaginary part represents the decay rate. From the data in Table \ref{fig:tablea}, it can be seen that due to the positive effective potential, the imaginary parts of all frequencies are negative, indicating the presence of a stable black hole solution in the pseudo-isothermal dark matter halo. Additionally, the imaginary parts of the frequencies represent the decay rate of oscillations, which is related to the dynamic evolution of QNMs. In the case of high overtone numbers, the imaginary part of QNM frequency decreases, which is consistent with the conclusion in the ordinary case.

\begin{figure*}[b]
	\begin{tabular}{ccc}
		\begin{minipage}{0.5\hsize}
			\centering
			\includegraphics[width=7.2cm,height=4.7cm]{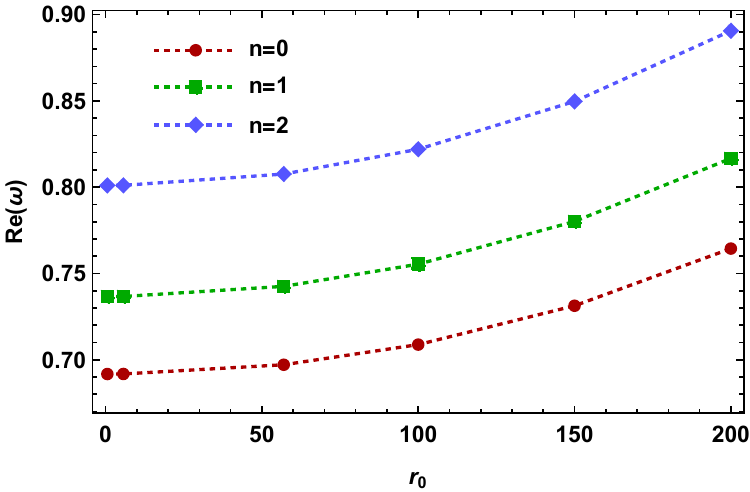}
		\end{minipage} &
		\begin{minipage}{0.5\hsize}
			\centering
			\includegraphics[width=7.2cm,height=4.7cm]{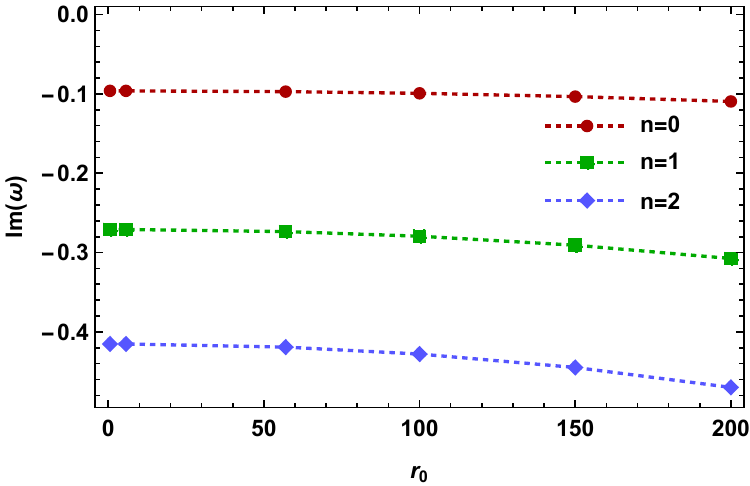}
		\end{minipage} 
	\end{tabular}
	\caption{The quasinormal modes of the black holes in scalar field for the state $l = 3$ in pseudo-isothermal halo model, where we fix the $\rho_0=0.0464 \, M_{\odot} / \mathrm{kpc}^3$.}
	\label{fig:QNMa}
	\begin{tabular}{ccc}
		\begin{minipage}{0.5\hsize}
			\centering
			\includegraphics[width=7.2cm,height=4.8cm]{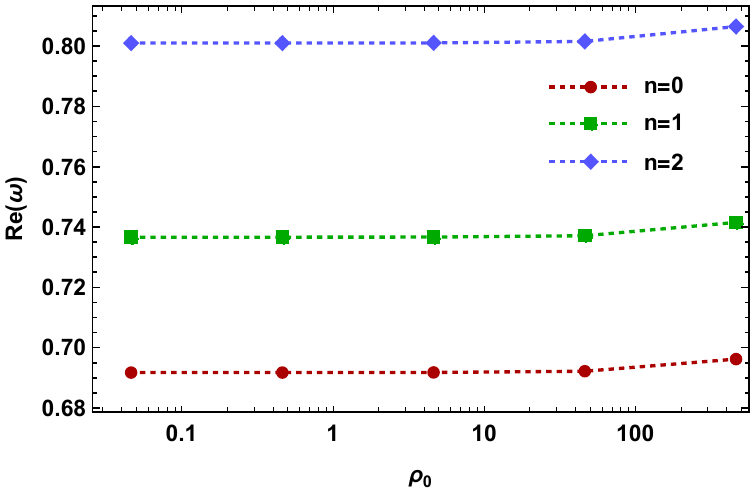}
		\end{minipage} &
		\begin{minipage}{0.5\hsize}
			\centering
			\includegraphics[width=7.2cm,height=4.8cm]{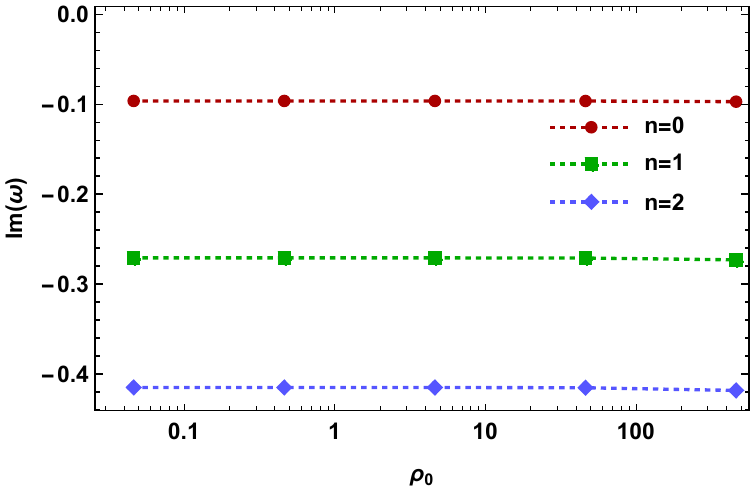}
		\end{minipage} 
	\end{tabular}
	\caption{The quasinormal modes of the black holes in scalar field for the state $l = 3$ in pseudo-isothermal halo model, where we fix the $r_0=0.57\, \mathrm{kpc}$. }
	\label{fig:QNMb}
\end{figure*}

The dynamic evolution of pseudo isothermal dark matter halos depend on core radius and density parameters. At first, by fixing density parameters $\rho_0=0.0464 \, M_{\odot} / \mathrm{kpc}^3$, the influence of core radius $r_0$ on the QNM frequencies $\omega$ can be demonstrated in Fig.\ref{fig:QNMa}. It is evident that \( \text{Re}(\omega) \) increases with the growth of \( r_0 \), while \( \text{Im}(\omega) \) decreases with the increase of \( r_0 \). This indicates that an increase in $r_0$ leads to the augmentation in the oscillation frequency, but accelerates the rate of decay. Then, by fixing core radius $r_0=0.57\, \mathrm{kpc}$, we study the influence of \( \rho_0 \) variations on QNMs, as shown in Fig.\ref{fig:QNMb}. By using the logarithmic-linear graph, we find the change in the imaginary part of the quasinormal frequency is still negligible, while the real part has a slight rise. This suggests that increasing $\rho_0$ does not significantly affect the energy dissipation, but could improve the oscillation performance of the system. 
\subsection{Connection between shadow and quasinormal modes}
In the eikonal limit, black hole shadow has a specific relationship with QNM frequency ($l\gg1$). This groundbreaking work is first proposed by Cardoso et al~\cite{Cardoso:2008bp}. Their research suggests that the real part of quasinormal models is related to the angular velocity $\Omega_{r_{ph}}$ of unstable circular orbits, while the imaginary part is related to the Lyapunov exponent  $\lambda_L$ determining the unstable timescale of orbits. Quasinormal frequencies can be obtained through the radius of the photon sphere via such connections,
\begin{align}\label{conneq}
	\omega_{l\gg1}=\Omega_{r_{ph}}l-i\Big(n+\frac{1}{2}\Big)|\lambda_L|.
\end{align}
Here $n$ and $l$ respectively represent the overtone number and the multiple number. The angular velocity $\Omega_{r_{ph}}$ at the unstable circular null geodesic and Lyapunov exponent $\lambda$ are given by the following expressions

\begin{equation}
	\begin{aligned}
		\Omega_{r_{ph}}=&\frac{\sqrt{f(r_{ph})}}{r_{ph}},\quad
		\lambda_L=&\sqrt{\frac{f(r_{ph})[2f(r_{ph})-r_{ph}^2f''(r_{ph})]}{2r_{ph}^2}}.
	\end{aligned}
\end{equation}
 The real and imaginary parts of the QNM frequency can be expressed as functions of the radius of the photon sphere. By utilizing the relationship between the photon sphere radius $r_{ph}$ and the shadow $R_s$, We can naturally obtain that the real part of the quasinormal modes is related to the shadow in the eikonal limit~\cite{Jusufi:2019ltj, Liu:2020ola},
\begin{align}
	\omega_{\mathfrak{R}}=\lim_{l\gg1}\frac{l}{R_s}.
\end{align}
Here $\omega_{\mathfrak{R}}$ represents the real part of QNMs and $R_s$ represents the black hole shadow. Therefore, Eq.(\ref{conneq}) can be rewritten as
\begin{align}
	\omega_{l\gg1}=\frac{l}{R_s}-i\Big(n+\frac{1}{2}\Big)|\lambda_L|.
\end{align}
The importance of this dependency is based on the fact that the black hole shadow can be directly obtained through astronomical observation. Therefore, it is viable to use the shadow instead of the angular velocity to represent the real part of the quasinormal modes. Using this method we can obtain the size of the black hole shadow from the real part of QNMs without using the geodesic method. The results of calculations using both the eikonal limit and the WKB methods are presented in Table \ref{fig:tableew}. By calculating and comparing the quasinormal frequencies for different multipole numbers, we have verified the effectiveness and accuracy of the WKB method. As the multipole numbers increases, the difference in both the real and imaginary parts of the frequency gradually decreases. This indicates that the eikonal limit method is highly accurate at large \( l \) values, with results very close to the WKB approximate. This is of great practical significance because the eikonal limit method is generally simpler for analytical calculations. High-precision results can provide reliable data support for studying the stability and oscillatory characteristics of black holes and other compact objects.

\begin{table}[t]
	\centering
	\setlength{\tabcolsep}{3pt} 
	\renewcommand{\arraystretch}{0.8} 
	\begin{tabular}{|c|c|c|c|c|}
		\hline
		&eikonal & WKB&$\Delta_{Re (\omega)}$ &$\Delta_{Im (\omega)}$ \\
		\hline
		$l=110$ & 21.169523170$-0.096225125i$ & 21.266328813$-0.096225016i$ & 0.455\% & -0.000114\% \\
		\hline
		$l=120$ & 23.094025276$-0.096225125i$ & 23.190782743$-0.096225033i$ & 0.417\% & -0.0000956\% \\
		\hline
		$l=130$ & 25.018527382$-0.096225125i$ & 25.115244056$-0.096225047i$ & 0.385\% & -0.0000815\% \\
		\hline
		$l=140$ & 26.943029489$-0.096225125i$ & 27.039711175$-0.096225058i$ & 0.358\% & -0.0000703\% \\
		\hline
		$l=150$ & 28.867531595$-0.096225125i$ & 28.964182944$-0.096225066i$ & 0.334\% & -0.0000613\% \\
		\hline
		$l=160$ & 30.792033701$-0.096225125i$ & 30.888658494$-0.096225074i$ & 0.313\% & -0.0000539\% \\
		\hline
		$l=170$ & 32.716535807$-0.096225125i$ & 32.813137158$-0.096225079i$ & 0.294\% & -0.0000478\% \\
		\hline
		$l=180$ & 34.641037914$-0.096225125i$ & 34.737618420$-0.096225084i$ & 0.278\% & -0.0000426\% \\
		\hline
		$l=190$ & 36.565540020$-0.096225125i$ & 36.662101870$-0.096225089i$ & 0.263\% & -0.0000383\% \\
		\hline
		$l=200$ & 38.490042126$-0.096225125i$ & 38.586587181$-0.096225092i$ & 0.25\% & -0.0000345\% \\
		\hline
	\end{tabular}
	\caption{We calculated the QNMs using both the eikonal limit and the WKB method, and computed the difference between the two method. The calculations were performed for \( l \) values ranging from $110$ to $200$.}
	\label{fig:tableew}
\end{table}
\subsection{Time evolution}
To investigate the time evolution of scalar field perturbations within dark matter halos hosting black holes, we opt for a time domain method developed by Gundlach, Price, and Pullin~\cite{Gundlach:1993tp}. By employing a time domain integration method to compute the evolution of $\Phi(t,r_*)$ at the fixed point $r_*$, we derive the time-domain profile. For the convenience of subsequent calculations, we re-write the wave-like Eq.(\ref{waveeq})
\begin{align}
	\frac{\partial^2\Phi}{\partial t^2}-\frac{\partial^2\Phi}{\partial r_{*}^2}+V(t,r_*)\Phi=0.
\end{align}
The finite difference method can be effectively utilized in handling time-domain integration. We will rewrite the above equation using light-cone coordinates, $u=t-r_*$ and $v=t+r_*$
\begin{align}\label{wave1}
	\Bigg(4\frac{\partial^2}{\partial u\partial v}+V(u,v)\Bigg)\Phi(u,v)=0.
\end{align}
To effectively integrate the above equation, we can employ a discretized integration approach~\cite{Nollert:1999ji, Moderski:2001gt, Moderski:2001tk},
\begin{align}
	\Phi(N)=\Phi(W)+\Phi(E)-\Phi(S)-h^2\frac{V(W)\Phi(W)+V(E)\Phi(E)}{8}+O(h^4).
\end{align}

Here, we use the following names for points in the \( u-v \) plane with a step size of \( h \): $N=(u+h,v+h)$,$W=(u+h,v)$,$E=(u,v+h)$ and $S=(u,v)$.  This method enables efficient numerical integration of the equation, offering accurate and effective results. Without loss of generality, we select  the Gaussian wave packet for two null surfaces as our initial conditions\cite{Gundlach:1993tp}, which are $u=u_0$ and $v=v_0$

\begin{figure*}[t]
	\begin{tabular}{ccc}
		\begin{minipage}{0.33\hsize}
			\centering
			\includegraphics[width=5.4cm,height=3.5cm]{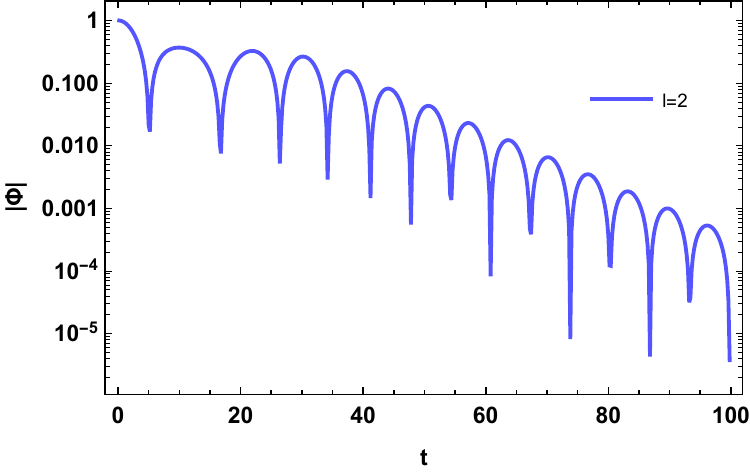}
		\end{minipage} &
		\begin{minipage}{0.33\hsize}
			\centering
			\includegraphics[width=5.4cm,height=3.5cm]{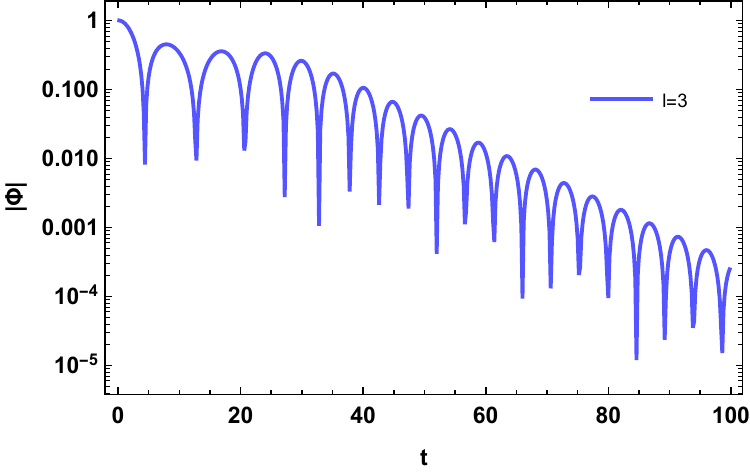}
		\end{minipage} &
		\begin{minipage}{0.33\hsize}
			\centering
			\includegraphics[width=5.4cm,height=3.5cm]{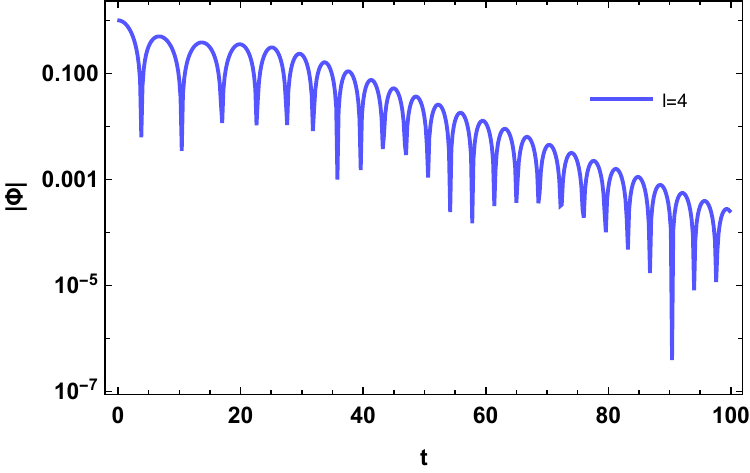}
		\end{minipage} 
	\end{tabular}
	\caption{The behavior of time-domain for scalar perturbation of black hole surrounded by pseudo-isothermal halo for different values of the parameter $l=2, 3, 4$.($r_0=0.57 \, \mathrm{kpc}$ and $\rho_0=0.0464 \, M_{\odot} /\mathrm{kpc}^3$)}
	\label{fig:Ptimea}
\end{figure*}
\begin{equation}
\begin{aligned}
	\Phi(u=u_0,v)&=A \, \exp\Bigg[-\frac{(v-v_c)^2}{2\sigma^2}\Bigg],\\
	\Phi(u,v=v_0)&=0,
\end{aligned}
\end{equation}
where we use $A=1$, $v_c=10$, $\sigma=3$ and $r_*=10$. We consider establishing a grid to cover the range of $u\in [-10,  90]$ and $v \in [10,  110]$, and set the grid factor $h = 0.2$. The evolution of QNMs is not influenced by the initial conditions or the small step size $h$. In this way, we can compute the time evolution of a black hole under the scale perturbation.

\begin{figure*}[b]
		\begin{tabular}{cc}
		\begin{minipage}{0.5\hsize}
			\centering
			\includegraphics[width=7.4cm,height=4.8cm]{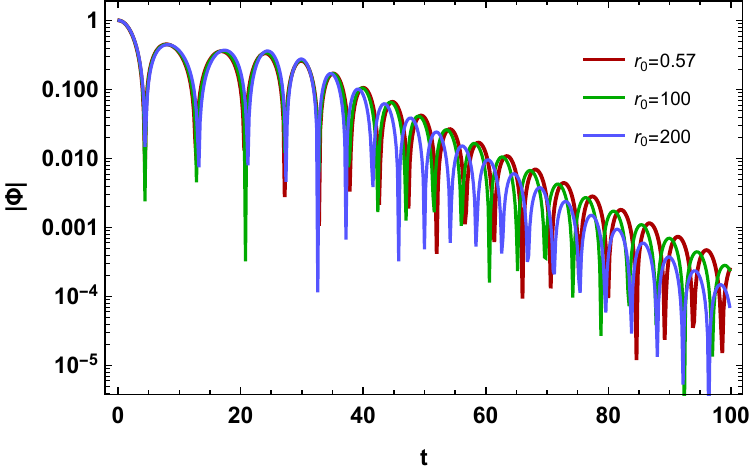}
		\end{minipage}&
		\begin{minipage}{0.5\hsize}
			\centering
			\includegraphics[width=7.4cm,height=4.8cm]{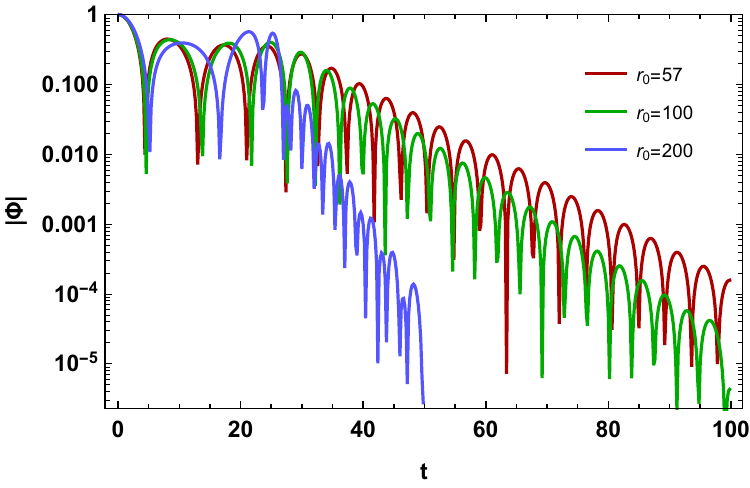}
		\end{minipage}
	\end{tabular}
	\caption{Semilogarithmic graph of the time-domain profiles for scalar perturbation of black hole surrounded by pseudo-isothermal halo  ($l=3$). In the left image, we set the parameter $\rho_0=0.0464 \, M_{\odot} / \mathrm{kpc}^3$, and in the right image, the parameter $\rho_0=0.464 \, M_{\odot} / \mathrm{kpc}^3$.}
	\label{fig:Ptimeb}
	
	\begin{tabular}{cc}
		\begin{minipage}{0.5\hsize}
			\centering
			\includegraphics[width=7.4cm,height=4.8cm]{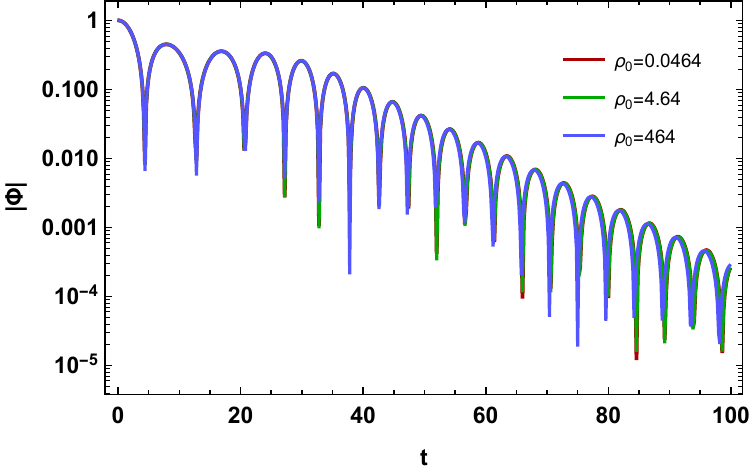}
		\end{minipage}&
		\begin{minipage}{0.5\hsize}
			\centering
			\includegraphics[width=7.4cm,height=4.8cm]{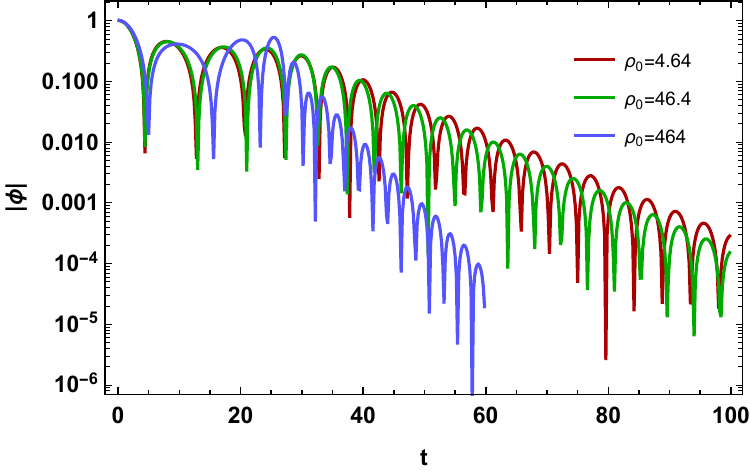}
		\end{minipage}
	\end{tabular}
	\caption{Semilogarithmic graph of the time-domain profiles for scalar perturbation of black hole surrounded by pseudo-isothermal halo  ($l=3$). In the left image, we set the parameter $r_0=0.57 \,\mathrm{kpc}$, and in the right image, the parameter $r_0=5.7 \,\mathrm{kpc}$.}
	\label{fig:Ptimec}
\end{figure*}
The QNMs ringing illustrates the time evolution of black hole spacetime of scalar perturbation. At first, by fixing dark matter parameters $r_0=0.57 \, \mathrm{kpc}$ and $\rho_0=0.0464 \, M_{\odot} / \mathrm{kpc}^3$ , we can obtain the result of multipole number for time evolution in Fig.\ref{fig:Ptimea}. It's shown that the growth of multipole numbers $l$ can increase the oscillation,  frequency without significantly increasing the damping time. We fix $l=3$  to study the effect of core radius on the time domain evolution in Fig.\ref*{fig:Ptimeb}. The central density $ \rho_0$ is set to $0.0464 \, M_{\odot} / \mathrm{kpc}^3$ in the left panel, and in the right panel, it is set to $\rho_0=0.464 \, M_{\odot} / \mathrm{kpc}^3$. The change of the time evolution graph is more significant when the density parameter is larger.  It is mainly reflected that the increase of core radius accelerates the damping of evolution, and also increases the oscillation frequency. Additionally, the influence of central density is studied by fixing $r_0=0.57 \,\mathrm{kpc}$ and $r_0=5.7 \,\mathrm{kpc}$ respectively, as shown in Fig.\ref{fig:Ptimec}. As discussed earlier, increasing $\rho_0$ leads to a slight increase in the real parts of the frequency. In the time domain, this is manifested by the oscillation frequency remaining nearly constant. This finding suggests that the time evolution of QNMs is minimally affected by the central density on a small scale. When the radius of the halo core reaches $5.7 \,\mathrm{kpc}$, the evolution profiles undergo significant changes throughout the entire period, and the decay rate of oscillations accelerates with increasing parameters. 
Increasing either $r_0 $ or $\rho_0$ accelerates the damping rate of oscillations. Larger values of $r_0$ and $\rho_0$ imply faster energy dissipation, which may be related to the density and distribution of the medium around the black hole. Variations in these parameters affect the propagation characteristics of gravitational waves, thereby influencing the signals observed on detectors. Studying the result of these parameters on perturbation evolution aids in understanding the physical environment around black holes and the propagation mechanisms of gravitational waves.

\section{Conclusions}\label{sect4}
In this paper, we analyzed the shadows and quasinormal modes (QNMs) of Schwarzschild-like black holes surrounded by pseudo-isothermal dark matter halos. We demonstrated that such black holes possess an unstable circular null geodesic on the equatorial plane outside the event horizon, indicating that the size of the black hole shadow can be approximated by the radius of the photon sphere. Our main focus was on the influence of dark matter halo parameters on the QNMs and the time evolution of scalar perturbations. Using the sixth-order WKB approximation and the finite difference method, we computed the QNM frequencies and the time-domain profiles.

The pseudo-isothermal dark matter halos are characterized by the halo core radius $r_0$ and the central halo density $\rho_0$. 
Our results showed that both the real and imaginary parts of the QNM frequencies increase as these parameters rise. We also examined the relationship between the quasinormal spectrum and time dependence in scattering phenomena.
It shows that increasing the multipole numbers $l$ increases the oscillation frequency, while the decay rate remains unchanged. Furthermore, larger dark matter parameters increase both the oscillation frequency and decay rate, with the impact of dark matter extending from the initial to the global phases of the ringdown.
Since astrophysical black holes have non-zero angular momentum, our work can be extended to kerr black holes.
Investigating the interaction between dark matter and black holes could provide valuable insights into the elusive nature of dark matter, potentially offering new clues through astrophysical observations.

\begin{acknowledgments}
We are grateful to Peng Wang, Yiqian Chen, and  Guangzhou Guo for useful discussions. This work is supported by NSFC Grant Nos. 12175212, 12275183, 12275184. This work is finished on the server from Kun-Lun in the Center for Theoretical Physics, School of Physics, Sichuan University.
\end{acknowledgments}


\begin{thebibliography}{99}
\bibitem{EventHorizonTelescope:2019dse}
Akiyama, Kazunori and others, First M87 Event Horizon Telescope Results. I. The Shadow of the Supermassive Black Hole, Astrophys. J. Lett. \textbf{875}, L1 (2019).

	
\bibitem{EventHorizonTelescope:2019ggy}
K.~Akiyama \textit{et al.} [Event Horizon Telescope],
First M87 Event Horizon Telescope Results. VI. The Shadow and Mass of the Central Black Hole,
Astrophys. J. Lett. \textbf{875}, no.1, L6 (2019). 

\bibitem{Ling:2021vgk}
R.~Ling, H.~Guo, H.~Liu, X.~M.~Kuang and B.~Wang,
Shadow and near-horizon characteristics of the acoustic charged black hole in curved spacetime,
Phys. Rev. D \textbf{104}, no.10, 104003 (2021).

\bibitem{Yang:2022btw}
J.~Yang, C.~Zhang and Y.~Ma,
Shadow and stability of quantum-corrected black holes,
Eur. Phys. J. C \textbf{83}, no.7, 619 (2023).


\bibitem{Haroon:2018ryd}
S.~Haroon, M.~Jamil, K.~Jusufi, K.~Lin and R.~B.~Mann,
Shadow and Deflection Angle of Rotating Black Holes in Perfect Fluid Dark Matter with a Cosmological Constant,
Phys. Rev. D \textbf{99}, no.4, 044015 (2019).



\bibitem{Vagnozzi:2022moj}
S.~Vagnozzi, R.~Roy, Y.~D.~Tsai, L.~Visinelli, M.~Afrin, A.~Allahyari, P.~Bambhaniya, D.~Dey, S.~G.~Ghosh and P.~S.~Joshi, \textit{et al.}
Horizon-scale tests of gravity theories and fundamental physics from the Event Horizon Telescope image of Sagittarius A,
Class. Quant. Grav. \textbf{40}, no.16, 165007 (2023).


  \bibitem{Jusufi:2019nrn}
K.~Jusufi, M.~Jamil, P.~Salucci, T.~Zhu and S.~Haroon,
Black Hole Surrounded by a Dark Matter Halo in the M87 Galactic Center and its Identification with Shadow Images,
Phys. Rev. D \textbf{100}, no.4, 044012 (2019).

\bibitem{Bambi:2019tjh}
C.~Bambi, K.~Freese, S.~Vagnozzi and L.~Visinelli,
Testing the rotational nature of the supermassive object M87* from the circularity and size of its first image, Phys. Rev. D \textbf{100}, no.4, 044057 (2019).

\bibitem{Vagnozzi:2019apd}
S.~Vagnozzi and L.~Visinelli,
Hunting for extra dimensions in the shadow of M87*,
Phys. Rev. D \textbf{100}, no.2, 024020 (2019).


\bibitem{Kumar:2020yem}
R.~Kumar, A.~Kumar and S.~G.~Ghosh,
Testing Rotating Regular Metrics as Candidates for Astrophysical Black Holes,
Astrophys. J. \textbf{896}, no.1, 89 (2020).

\bibitem{Ghosh:2020ece}
S.~G.~Ghosh, M.~Amir and S.~D.~Maharaj,
Ergosphere and shadow of a rotating regular black hole,
Nucl. Phys. B \textbf{957}, 115088 (2020).


\bibitem{Adler:2022qtb}
S.~L.~Adler and K.~S.~Virbhadra,
Cosmological constant corrections to the photon sphere and black hole shadow radii,
Gen. Rel. Grav. \textbf{54}, no.8, 93 (2022).


	
  \bibitem{Konoplya:2011qq}
R.~A.~Konoplya and A.~Zhidenko,Quasinormal modes of black holes: From astrophysics to string theory,
Rev. Mod. Phys. \textbf{83}, 793-836 (2011).

  \bibitem{Vishveshwara:1970zz}
C.~V.~Vishveshwara, Scattering of Gravitational Radiation by a Schwarzschild Black-hole, Nature \textbf{227}, 936-938 (1970).

\bibitem{Sun:2023woa}
Q.~Sun, Q.~Li, Y.~Zhang and Q.~Q.~Li,
Quasinormal modes, Hawking radiation and absorption of the massless scalar field for Bardeen black hole surrounded by perfect fluid dark matter,
Mod. Phys. Lett. A \textbf{38}, no.22n23, 2350102 (2023).

\bibitem{Pedrotti:2024znu}
D.~Pedrotti and S.~Vagnozzi,
Quasinormal modes-shadow correspondence for rotating regular black holes,
[arXiv:2404.07589 [gr-qc]].


\bibitem{Zhang:2021bdr}
C.~Zhang, T.~Zhu and A.~Wang,
Gravitational axial perturbations of Schwarzschild-like black holes in dark matter halos,
Phys. Rev. D \textbf{104}, no.12, 124082 (2021).

\bibitem{Hendi:2020zyw}
S.~H.~Hendi, A.~Nemati, K.~Lin and M.~Jamil,
Instability and phase transitions of a rotating black hole in the presence of perfect fluid dark matter, Eur. Phys. J. C \textbf{80}, no.4, 296 (2020).

\bibitem{Jafarzade:2020ova}
K.~Jafarzade, M.~Kord Zangeneh and F.~S.~N.~Lobo,
Shadow, deflection angle and quasinormal modes of Born-Infeld charged black holes,
JCAP \textbf{04}, 008 (2021).



\bibitem{Anacleto:2021qoe}
M.~A.~Anacleto, J.~A.~V.~Campos, F.~A.~Brito and E.~Passos,
Quasinormal modes and shadow of a Schwarzschild black hole with GUP,
Annals Phys. \textbf{434}, 168662 (2021).

\bibitem{Ghosh:2022gka}
R.~Ghosh, M.~Rahman and A.~K.~Mishra,
Regularized stable Kerr black hole: cosmic censorships, shadow and quasi-normal modes,
Eur. Phys. J. C \textbf{83}, no.1, 91 (2023).

\bibitem{Ghosh:2023etd}
R.~Ghosh, N.~Franchini, S.~H.~V\"olkel and E.~Barausse,
Quasinormal modes of nonseparable perturbation equations: The scalar non-Kerr case,
Phys. Rev. D \textbf{108}, no.2, 024038 (2023).

\bibitem{Hamil:2024ppj}
B.~Hamil and B.~C.~L\"utf\"uo\u{g}lu, Noncommutative Schwarzschild black hole surrounded by quintessence: Thermodynamics, Shadows and Quasinormal modes, Phys. Dark Univ. \textbf{44}, 101484 (2024).

\bibitem{Lambiase:2023hng}
G.~Lambiase, R.~C.~Pantig, D.~J.~Gogoi and A.~\"Ovg\"un,
Investigating the connection between generalized uncertainty principle and asymptotically safe gravity in black hole signatures through shadow and quasinormal modes,
Eur. Phys. J. C \textbf{83}, no.7, 679 (2023) .



\bibitem{Jusufi:2019ltj}
K.~Jusufi,
Quasinormal Modes of Black Holes Surrounded by Dark Matter and Their Connection with the Shadow Radius,
Phys. Rev. D \textbf{101}, no.8, 084055 (2020).


\bibitem{Schutz:1985km}
B.~F.~Schutz and C.~M.~Will, BLACK HOLE NORMAL MODES: A SEMIANALYTIC APPROACH, Astrophys. J. Lett. \textbf{291}, L33-L36 (1985).

\bibitem{Iyer:1986np}
S.~Iyer and C.~M.~Will, Black Hole Normal Modes: A {WKB} Approach. 1. Foundations and Application of a Higher Order {WKB} Analysis of Potential Barrier Scattering, Phys. Rev. D \textbf{35}, 3621 (1987).


\bibitem{Ferrari:1984zz}
V.~Ferrari and B.~Mashhoon, New approach to the quasinormal modes of a black hole, Phys. Rev. D \textbf{30}, 295-304 (1984).

\bibitem{Churilova:2021nnc}
M.~S.~Churilova, R.~A.~Konoplya and A.~Zhidenko, Analytic formula for quasinormal modes in the near-extreme Kerr-Newman\textendash{}de Sitter spacetime governed by a non-P\"oschl-Teller potential,
Phys. Rev. D \textbf{105}, no.8, 084003 (2022) .

\bibitem{Leaver:1985ax}
E.~W.~Leaver, An Analytic representation for the quasi normal modes of Kerr black holes, Proc. Roy. Soc. Lond. A \textbf{402}, 285-298 (1985).

\bibitem{Dolan:2007mj}
S.~R.~Dolan, Instability of the massive Klein-Gordon field on the Kerr spacetime, Phys. Rev. D \textbf{76}, 084001 (2007).


\bibitem{Aso:2013eba}
Y.~Aso \textit{et al.} [KAGRA],
Interferometer design of the KAGRA gravitational wave detector,
Phys. Rev. D \textbf{88}, no.4, 043007 (2013).

\bibitem{Liu:2020eko}
S.~Liu, Y.~M.~Hu, J.~d.~Zhang and J.~Mei,
Science with the TianQin observatory: Preliminary results on stellar-mass binary black holes,
Phys. Rev. D \textbf{101}, no.10, 103027 (2020).

\bibitem{Ruan:2018tsw}
W.~H.~Ruan, Z.~K.~Guo, R.~G.~Cai and Y.~Z.~Zhang,
Taiji program: Gravitational-wave sources,
Int. J. Mod. Phys. A \textbf{35}, no.17, 2050075 (2020).


 
  \bibitem{deSwart:2017heh}
 J.~de Swart, G.~Bertone and J.~van Dongen, How Dark Matter Came to Matter,
 Nature Astron. \textbf{1}, 0059 (2017).


\bibitem{Zwicky:1933gu}
F.~Zwicky,
Die Rotverschiebung von extragalaktischen Nebeln,
Helv. Phys. Acta \textbf{6}, 110-127 (1933).

  
 \bibitem{Navarro:1995iw}
 J.~F.~Navarro, C.~S.~Frenk and S.~D.~M.~White,
 The Structure of cold dark matter halos,
 Astrophys. J. \textbf{462}, 563-575 (1996).


\bibitem{Navarro:1996gj}
J.~F.~Navarro, C.~S.~Frenk and S.~D.~M.~White,
A Universal density profile from hierarchical clustering,
Astrophys. J. \textbf{490}, 493-508 (1997).

  

\bibitem{Robles:2012uy}
V.~H.~Robles and T.~Matos,
Flat Central Density Profile and Constant DM Surface Density in Galaxies from Scalar Field Dark Matter,
Mon. Not. Roy. Astron. Soc. \textbf{422}, 282-289 (2012)


\bibitem{McGaugh:1998tq}
S.~S.~McGaugh and W.~J.~G.~de Blok,
Testing the dark matter hypothesis with low surface brightness galaxies and other evidence, Astrophys. J. \textbf{499}, 41 (1998).

\bibitem{Spergel:1999mh}
D.~N.~Spergel and P.~J.~Steinhardt,
Observational evidence for selfinteracting cold dark matter,
Phys. Rev. Lett. \textbf{84}, 3760-3763 (2000).

\bibitem{Kamionkowski:1999vp}
M.~Kamionkowski and A.~R.~Liddle,
The Dearth of halo dwarf galaxies: Is there power on short scales?,
Phys. Rev. Lett. \textbf{84}, 4525-4528 (2000).


 
\bibitem{Sadeghian:2013laa}
L.~Sadeghian, F.~Ferrer and C.~M.~Will,
Dark matter distributions around massive black holes: A general relativistic analysis,
Phys. Rev. D \textbf{88}, no.6, 063522 (2013).

\bibitem{Fields:2014pia}
B.~D.~Fields, S.~L.~Shapiro and J.~Shelton,
Galactic Center Gamma-Ray Excess from Dark Matter Annihilation: Is There A Black Hole Spike?,
Phys. Rev. Lett. \textbf{113}, 151302 (2014).


\bibitem{Zhao:2023tyo}
Y.~Zhao, B.~Sun, K.~Lin and Z.~Cao,
Axial gravitational ringing of a spherically symmetric black hole surrounded by dark matter spike,
Phys. Rev. D \textbf{108}, no.2, 024070 (2023).


\bibitem{Begeman:1989kf}
K.~G.~Begeman,
H I rotation curves of spiral galaxies. I - NGC 3198,
Astron. Astrophys. \textbf{223}, 47-60 (1989).

  \bibitem{Yang:2023tip}
 Y.~Yang, D.~Liu, A.~\"Ovg\"un, G.~Lambiase and Z.~W.~Long, Black hole surrounded by the pseudo-isothermal dark matter halo, Eur. Phys. J. C \textbf{84} (2024) no.1, 63 (2024).
 
 \bibitem{Lin:2019yux}
 H.~N.~Lin and X.~Li, The Dark Matter Profiles in the Milky Way, Mon. Not. Roy. Astron. Soc. \textbf{487}, no.4, 5679-5684 (2019).
 

\bibitem{Johannsen:2013vgc}
T.~Johannsen,
Photon Rings around Kerr and Kerr-like Black Holes,
Astrophys. J. \textbf{777}, 170 (2013).

  
 
 \bibitem{Hioki:2009na}
 K.~Hioki and K.~i.~Maeda,
Measurement of the Kerr Spin Parameter by Observation of a Compact Object's Shadow,
 Phys. Rev. D \textbf{80}, 024042 (2009).

  
\bibitem{Cardoso:2008bp}
 V.~Cardoso, A.~S.~Miranda, E.~Berti, H.~Witek and V.~T.~Zanchin, Geodesic stability, Lyapunov exponents and quasinormal modes, Phys. Rev. D \textbf{79}, no.6, 064016 (2009).
 
 \bibitem{Liu:2020ola}
 C.~Liu, T.~Zhu, Q.~Wu, K.~Jusufi, M.~Jamil, M.~Azreg-A\"\i{}nou and A.~Wang,
 Shadow and quasinormal modes of a rotating loop quantum black hole,
 Phys. Rev. D \textbf{103}, no.8, 089902 (2021).
 
 
\bibitem{Gundlach:1993tp}
C.~Gundlach, R.~H.~Price and J.~Pullin,
Late time behavior of stellar collapse and explosions: 1. Linearized perturbations,
Phys. Rev. D \textbf{49}, 883-889 (1994).


\bibitem{Nollert:1999ji}
 H.~P.~Nollert,
TOPICAL REVIEW: Quasinormal modes: the characteristic `sound' of black holes and neutron stars,
 Class. Quant. Grav. \textbf{16}, R159-R216 (1999).


\bibitem{Moderski:2001gt}
R.~Moderski and M.~Rogatko,
Late time evolution of a charged massless scalar field in the space-time of a dilaton black hole,
Phys. Rev. D \textbf{63}, 084014 (2001).
  
 
 \bibitem{Moderski:2001tk}
 R.~Moderski and M.~Rogatko,
 Late time evolution of a selfinteracting scalar field in the space-time of dilaton black hole,
 Phys. Rev. D \textbf{64}, 044024 (2001).


\end{thebibliography}
\end{document}